\documentclass[prl.aps,twocolumn,showpacs,floats,superscriptaddress,hyperref=pdftex]{revtex4-1}
\usepackage{graphicx}
\usepackage{dcolumn}
\usepackage{bm}
\usepackage{color}
\usepackage{hyperref}
\hypersetup{colorlinks=true,linkcolor=blue,anchorcolor=blue,citecolor=blue,filecolor=blue,urlcolor=blue,bookmarksnumbered=true,pdfview=FitB}
\begin{document}


\author{Andrey S. Mishchenko}
\affiliation{RIKEN Center for Emergent Matter Science (CEMS), Wako, Saitama 351-0198, Japan}
\affiliation{National Research Center ``Kurchatov Institute'', 123182 Moscow, Russia}

\author{Lode Pollet}
\affiliation{Department of Physics, Arnold Sommerfeld Center for Theoretical Physics,
University of Munich, Theresienstrasse 37, 80333 Munich, Germany}

\author{Nikolay V. Prokof'ev}
\affiliation{Department of Physics, University of Massachusetts, Amherst, MA 01003, USA}
\affiliation{National Research Center ``Kurchatov Institute'', 123182 Moscow, Russia}

\author{Abhishek Kumar}
\affiliation{Department of Physics, University of Florida,
2001 Museum Road, Gainesville, FL 32611, USA}

\author{Dmitrii L. Maslov}
\affiliation{Department of Physics, University of Florida,
2001 Museum Road, Gainesville, FL 32611, USA}

\author{Naoto Nagaosa}
\affiliation{RIKEN Center for Emergent Matter Science (CEMS), Wako, Saitama 351-0198, Japan}
\affiliation{Department of Applied Physics, The University of Tokyo, 7-3-1 Hongo, Bunkyo-ku, Tokyo 113, Japan}

\title{Polaron mobility in the ``beyond quasiparticles" regime}

\begin{abstract}

In a number of physical situations, from polarons to Dirac liquids and to non-Fermi liquids, one encounters the ``beyond quasiparticles" regime, in which the inelastic scattering rate exceeds the thermal energy of quasiparticles. Transport in this regime cannot be described by the kinetic equation. We employ the Diagrammatic Monte Carlo method to study the mobility of a Fr\"{o}hlich polaron in this regime and discover a number of non-perturbative effects: a strong violation of the Mott-Ioffe-Regel criterion at intermediate and strong couplings,  a mobility minimum at $T \sim \Omega$ in the strong-coupling limit ($\Omega$ is the optical mode frequency), a substantial delay in the onset of an exponential dependence of the mobility  for $T<\Omega$ at intermediate coupling, and complete smearing of the Drude peak at strong coupling. These effects should be taken into account when interpreting mobility data in materials with strong electron-phonon coupling.
\end{abstract}

\maketitle


Mobility of non-degenerate charge carriers in ionic semiconductors with strong
electron-phonon coupling is a long-standing problem \cite{Low,LK,Alexandrov}.
Textbook treatment starts with the notion of
a transport scattering time
$\tau$ controlling momentum relaxation in the
equation of motion $\dot{{\mathbf p}}=
{\mathbf E}- {\mathbf p}/\tau$,
where ${\mathbf E}$ is
the external electric field
  (we set $e=1$, $\hbar =1$, and $k_B=1$, for brevity). The underlying
 assumption
 is
 that all
 effects
originating from
coupling
the particle to
its environment (in this work we
consider
a
single particle
coupled to
a translationally-invariant
bath)
are reduced to an effective
friction force.
The
 equation of motion
  leads to
  a
  familiar result for
  the
  frequency-dependent Drude
   mobility,
\begin{equation}
\mu (\omega) = \frac{\tau/M}{1 -i \omega \tau } \;,
\label{Drude}
\end{equation}
with $M$
being
 the bare particle mass.

Deducing
the
scattering time from
the
DC mobility by using $\tau=\mu (0)M \equiv \mu M $ faces a problem
because coupling to the environment
necessarily leads to
mass
renormalization,
$M \to M^{*}$. One
might
argue that substituting $M^{*}$ instead of $M$ into Eq.~(\ref{Drude})
would
solve
the issue. However, this procedure implicitly assumes that $M^{*}$ can be measured
or calculated
separately in
a setup where
 particles propagate coherently
and
relaxation processes can be neglected at the relevant energy scales. In other words,
one has to require that $E \tau (E,T) \gg 1$ (in general, $\tau$ may depend on the
particle energy, $E$), or
\begin{equation}
\tau (T) \gg 1/T \;,
\label{IR}
\end{equation}
if $E$ is replaced with the typical thermal energy $ (d/2)T$, with $d$
being
 the spatial dimension.
Here,  $\tau(T)$ should be understood as the {\em inelastic} scattering time. For elastic scattering, either by impurities or phonons above the Bloch-Gr{\"u}neisen temperature, condition (\ref{IR}) is not relevant \cite{physkin}. 
Likewise, it can be strongly violated in lattice models in the hopping mobility regime \cite{Holstein59,minimum}
when the local rather than momentum representation is more adequate.  
Condition (\ref{IR}) can be reformulated as $\ell \gg \lambda_T $, where
$\ell$ is the mean free path and $\lambda_T$ is the particle de Broglie wavelength.
In this form, it coincides with the ``thermal'' version of the
Mott-Ioffe-Regel (MIR)
criterion for the validity of the kinetic equation approach, which has attracted a lot of attention recently
in the context of bad and strange metals \cite{hartnoll_book,hartnoll,bruin}.

In this Letter, we analyze the violation of the MIR criterion for a tractable yet non-trivial system, namely the Fr{\"o}hlich polaron model~\cite{Frohlich}.
Apart from being a canonical polaron problem \cite{Landau,Appel},
it plays the main role in understanding
charge transport
in ionic semiconductors. It is described by the Hamiltonian $H=H_{\rm e}+H_{\rm ph}+H_{\rm e-ph}$ where
\begin{equation}
H_{\rm e} = \sum_{\mathbf k} \frac{k^2}{2M} \, c_{\mathbf k}^{\dagger }  c_{\mathbf k}^{\,} \;, \;\;\;\;
H_{\rm ph} = \sum_{\mathbf q} \Omega \, b_{\mathbf q}^{\dagger }  b_{\mathbf q}^{\,},
\label{H0}
\end{equation}
\begin{equation}
H_{\rm e-ph} = \sum_{{\mathbf k},{\mathbf q}} V({\mathbf q}) \, (b_{\mathbf q}^{\dagger } - b_{-\mathbf q}^{\,})
\, c_{{\mathbf k}-{\mathbf q}}^{\dagger }  c_{\mathbf k}^{\,} \;,
\label{Heph}
\end{equation}
with
$V({\mathbf q})= i \sqrt{2^{3/2} \pi \Omega^{3/2}\alpha/M^{1/2}q^2}$.
Coupling between electrons and optical phonons with energy $\Omega$ can no longer
be assumed weak when the dimensionless constant $\alpha >1$.
Perturbation theory predicts that the polaron $Z$-factor in the ground state is
given by $Z = 1 - \alpha/2$ to leading order.
As
revealed by stochastically exact Diagrammatic Monte Carlo (diagMC) simulations~\cite{us1,us2},
the actual
$Z$-factor is
about
$0.3$ already at $\alpha=2$
 and
 drops down
to
$\approx 0.01$ at $\alpha=6$.

As far as the Fr\"{o}hlich polaron mobility at finite temperature is concerned,
rigorous analytical results can only be obtained within
some version of  perturbation theory, based either on weak coupling or the Migdal theorem.
In the anti-adiabatic regime
($T\ll \Omega$), one can only rely on weak coupling
($\alpha \ll 1$). This  limit was considered by Kadanoff and Langreth \cite{LK},
who
calculated
 all diagrams for $\mu$ up to order ${\cal O}(1/\alpha)+{\cal O}(1)$
with the result
\begin{equation}
\mu=\frac{e^{\Omega/T}}{2M^{*}\alpha \Omega}
= \frac{1/\alpha -1/6}{2M
\Omega
}e^{\Omega/T}\, ,
\;\;\;(T\ll \Omega,\alpha\ll 1)\;,
\label{LK}
\end{equation}
The ${\cal O}(1)$ correction accounts for
effective mass renormalization, $M^{*}=M/(1-\alpha/6)$,
and is legitimate. There is also a number of variational results;
a widely used one is the Low and Pines formula~\cite{Low},
which has the same form
as (\ref{LK}) with $M^*=M(1+\alpha/6)^3/f(\alpha)$,
where $f(\alpha)\approx 5/4$
for $3 < \alpha < 6$.
In the adiabatic limit ($T\gg \Omega$), one can neglect
vertex corrections when computing the scattering rate
and find the mobility from the kinetic equation, whose
solution is radically simplified by the fact that scattering
is quasi-elastic. The result is a mobility
that slowly decreases with $T$~\cite{Supplemental},
\begin{equation}
\mu=\frac{4}{3 \sqrt{\pi} \alpha M \sqrt{\Omega T}}\; \;\;\;\; (T\gg \Omega) \,.
\label{Maslov}
\end{equation}
The  $1/\sqrt{T}$ scaling of $\mu$ is a combination of
the $T$ scaling
of
the
phonon occupation numbers
and the $\sqrt{T}$ scaling of the thermal velocity. In contrast to Eq.~(\ref{LK}), $\alpha$ is not
is not required to be small for Eq.~(\ref{Maslov}) to be valid.

For $\alpha \gg 1$ both limits predict that at $T\sim \Omega$
one enters
the ``beyond quasiparticles" regime,
in which
$\tau(T)$ --defined as $M\mu(T)$--
is shorter than the Planckian bound.
As far as we know, the mobility of Fr\"{o}hlich  polarons
has never been computed with controlled accuracy in this parameter regime.

In this Letter, we address this unsolved problem
by extending the  diagMC method \cite{us1,us2} to finite temperatures
in combination with analytic continuation methods \cite{us2,us3,tutorialtext}
and computing
the mobility from the Matsubara current-current correlation function. We find that
for $\alpha >1$ the MIR condition, formulated
as
 $\mu \gg 1/MT$, is indeed violated. Contrary to
expectations based on perturbation theory, the exponential increase
of the mobility with decreasing temperature,
Eq.~(\ref{LK}), is observed
only at temperatures significantly below $\Omega$. Even more surprising is
a mobility minimum developing at
$T \sim \Omega$ for large values of $\alpha$. We interpret both effects as a competition between the
decreasing number of thermal excitations and
a strong enhancement of the polaron mass, both of which taking place
as temperature is lowered.
\begin{figure}
\vspace{0.05cm}
\centering
\includegraphics[width=0.76\linewidth]{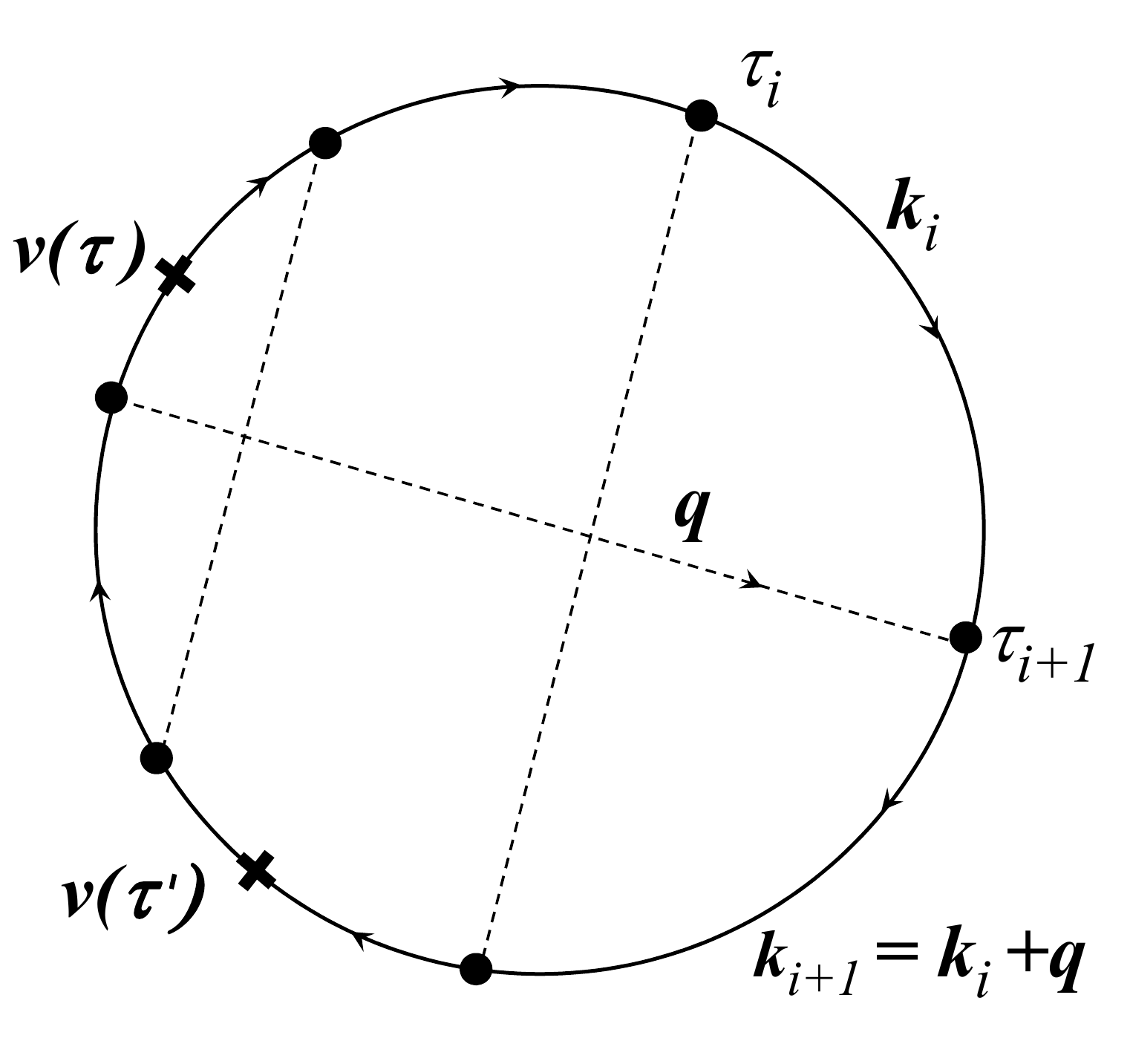}
\caption{\label{fig:1}
A typical Feynman diagram contributing to the current-current correlation function.
Solid (dashed) lines stand for electron (phonon) propagators $G$ ($D$), dots are
representing interaction vertices $\vert V \vert$, and crosses mark velocity
measurements.
}
\end{figure}

\noindent
{\it Method}.
As far as the diagMC method is concerned, our finite temperature calculation (in which we used the units $M=\Omega=1$)
is  similar to that employed for calculating the
optical conductivity
in Ref.~\cite{us4} for $T = \beta^{-1} \to 0$.
The
difference is that now phonon propagators in imaginary time are
temperature dependent,
$D(\tau) = [e^{-\Omega \tau} + e^{\Omega \tau -\Omega \beta }] / [1-e^{-\Omega \beta}]$,
and we do not assume that $e^{-\Omega \beta}$ is small. A typical diagram contributing
to the
current-current
correlation function
$C(\tau-\tau ') = d^{-1}\langle {\mathbf v}(\tau) {\mathbf v}(\tau ') \rangle
\equiv (dM^{2})^{-1} \langle {\mathbf k}(\tau) {\mathbf k}(\tau ') \rangle$
is shown in Fig.~\ref{fig:1}.
By using  Matsubara frequencies,  $\omega_m= 2\pi m T$ (with $m$ integer),
we obtain an efficient unbiased Monte Carlo estimator for its Fourier transform,
\begin{equation}
C_m = \int_0^\beta d\tau C(\tau) \, e^{i\omega_m \tau} \equiv \frac{1}{d \beta M^2}
\langle \vert {\mathbf k}_m \vert^2 \rangle.
\label{Cm}
\end{equation}
The value of ${\mathbf k}_m$ for a  diagram of order $N$ is
readily found from the electron momenta $\{ {\mathbf k}_i \}$
in imaginary time intervals
$i=1, \dots , 2N$ ranging from $\tau_i$ to $\tau_{i+1}$ (cf. Fig~\ref{fig:1}),
\begin{equation}
{\mathbf k}_m =\int_0^\beta d\tau {\mathbf k} (\tau) \, e^{i\omega_m \tau} =
\sum_{i}^{2N} {\mathbf k}_i
\frac{ e^{i\omega_m \tau_{i+1}} - e^{i\omega_m \tau_{i}}}{i\omega_m}\,.
\label{CmMC}
\end{equation}

There are several exact and asymptotic relations relations that the correlation function
has to satisfy. Any of them can be used
as
an independent check on
the
 accuracy
of
the
data.
For the equal-time correlator
we
find
that $T\sum_m C_m = \langle k^2 \rangle /dM^2$.
The sum rule for the mobility translates into $C_{m=0} = 1$. The asymptotic behavior in the limit
of large $|m|$ comes from diagrams containing very short time intervals
$\tau_{i+1} - \tau_{i}=\Delta \tau \to 0$ with large particle momenta
$k_i \sim 1/\sqrt{\Delta \tau } \to \infty$. Dressing such intervals by vertex corrections
is not necessary
because
it
would result in
additional factors of $ \sqrt{\Delta \tau }$.
Therefore, one can use
 perturbation theory to compute the
 high-frequency limit with the
final result
\begin{equation}
C_{m\to \infty} \to \frac{2\sqrt{2}\, \alpha \coth [\Omega \beta /2 ]}
{d \, \omega_m^{3/2}} \;.
\label{largem}
\end{equation}

To determine the mobility, one needs to perform the
analytic continuation numerically,
which amounts to inverting
the Kramers-Kronig transformation
\begin{equation}
C_{m} = \frac{2}{\pi} \int_{0}^{\infty} d\omega \,
\frac{ \omega^2 \mu (\omega)} {\omega^2 + \omega_m^2} \;.
\label{Cmmu}
\end{equation}
The asymptotic behavior implied by $C_m$ immediately
yields
the
asymptotic behavior of  $
\mu(\omega )$
\[
\mu (\omega \to \infty ) \to \frac{2 \alpha \coth [\Omega \beta /2 ]}
{d\,  \omega^{5/2}} \;,
\]
which can be used to subtract
the leading tail contribution when performing the
analytic continuation. The nature of Eq.~(\ref{Cmmu}) is such that even tiny error bars on $C_m$ translate into large uncertainties on integrals of $\mu (\omega )$ over
physically relevant intervals.
All errors on the $\mu (\omega )$ function itself
are conditional and depend on constraints for allowed behavior \cite{us3}.
In this work, Monte Carlo data for the lowest Matsubara frequencies were accurate at
the level of five to six significant digits.

Our analytic continuation method for solving Eq.~(\ref{Cmmu}) is similar to the one used
in Ref.~\cite{minimum}. The main difference is that we extract $\mu (\omega)$
from the
data
parameterized by the
Matsubara frequency
rather than imaginary time.
We also employ
more conservative protocols for computing
both
the mobility and its errorbars from multiple solutions of the
stochastic optimization with the consistent constraints method (see Supplemental material)~\cite{us2,us3}.
\begin{figure}
\vspace{0.05cm}
\centering
\includegraphics[width=0.9\linewidth]{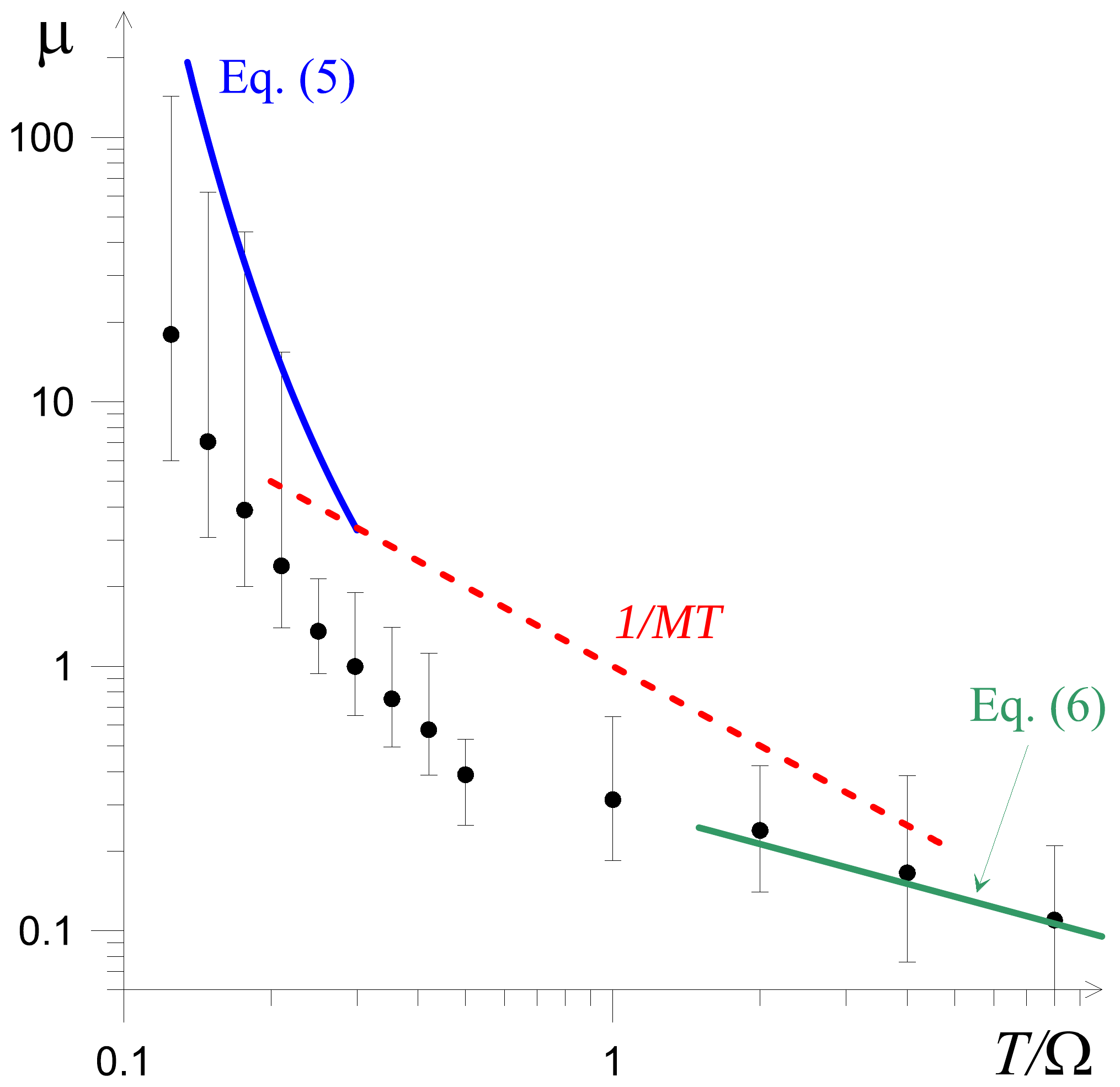}
\caption{\label{fig:2} (color online).
Black dots: mobility of
a
Fr\"{o}hlich polaron as a function of temperature at intermediate
coupling ($\alpha = 2.5$).
Blue and green lines are predictions based
on Eqs.~(\ref{LK}) and (\ref{Maslov}), respectively.
Below the dashed red line, $\mu =1/MT$, the MIR criterion is violated.
}
\end{figure}

\noindent
{\it Results}.
To begin with, we verified that our results for the mobility are consistent with the predictions
(\ref{LK}) and (\ref{Maslov}) based on perturbation theory for small $\alpha=0.25$.
In this case, the MIR criterion is not violated
even
at $T\sim \Omega$,
{\it i.e.}, the mobility remains large compared to $1/MT$.
When the coupling strength is increased to $\alpha=2.5$, the situation changes.
As Fig.~\ref{fig:2} shows, the MIR criterion  is violated over a broad temperature interval
$0.2< T/\Omega <10$.
The data for $T>\Omega$ approach the limiting form in Eq.~(\ref{Maslov}), as expected,
because Eq.~(\ref{Maslov}) is valid for any $\alpha$. On the other hand, one should not expect
the low-temperature formula (\ref{LK}) to be valid beyond weak coupling.
Indeed, it is clear from Fig.~\ref{fig:2} that the slow temperature dependence extends down to at least
$T\sim \Omega/2$.
At lower temperatures, our data are consistent with the exponential increase of the mobility,
but uncertainties amplified by the analytic
continuation procedure become too large for a meaningful analysis.
\begin{figure}
\vspace{0.1cm}
\centering
\includegraphics[width=0.9\linewidth]{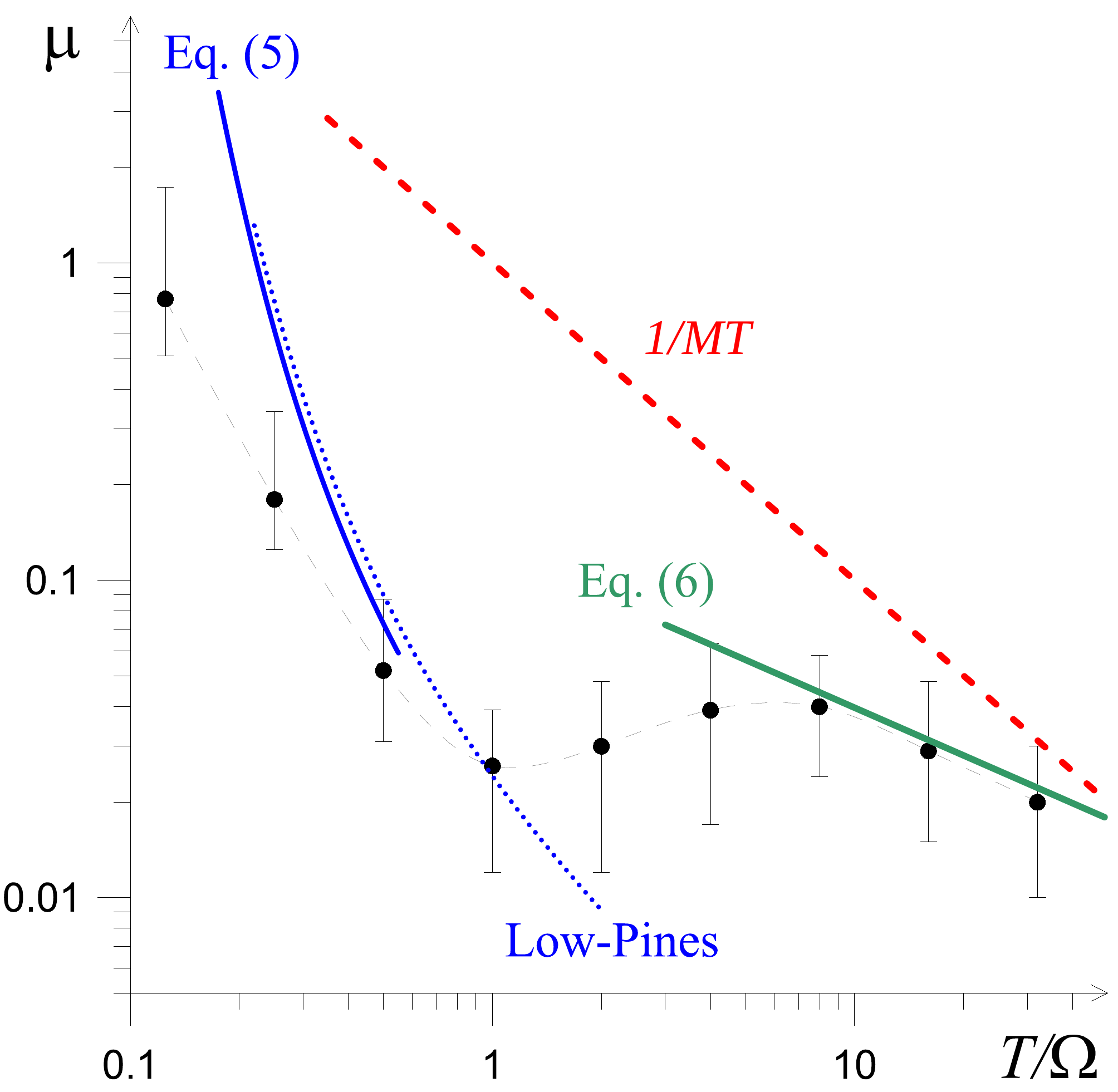}
\caption{\label{fig:3} (color online).
Mobility of a Fr\"{o}hlich polaron as a function of temperature at strong
coupling ($\alpha = 6$). All notations are identical to those
in Fig.~\ref{fig:2}. Eq.~(\ref{LK}) is plotted using the exact result $M^*=7.3$ instead of its perturbative approximation;
otherwise the curve would go unphysical. The Low-Pines formula
\cite{Low} is shown
by
a dotted line \cite{comment_LP}.
}
\end{figure}

A delay (at $T<\Omega$) in the onset of the exponential dependence may be anticipated already from perturbation theory. Indeed, matching Eqs.~(\ref{LK}) and (\ref{Maslov}), we find that the crossover temperature
\begin{equation}
T_{\text{cr}}\sim \frac{\Omega}{\ln \lambda}\left(1-\frac{\ln\ln\lambda}{2\ln\lambda}+\dots\right)
\end{equation}
with $\lambda\equiv M^*/M$
is
(logarithmically) suppressed compared to $\Omega$ for $\lambda\gg 1$.

One
might
argue that the MIR criterion cannot be specified down to a well-defined numerical
factor and, thus, having $\mu MT =1/4$ for $\alpha=2.5$ is merely a borderline situation.
However, the strong coupling case ($\alpha=6$) (cf.~Fig.~\ref{fig:3})
shows that there is no obvious limit on how small the value of $\mu MT$ can be. Even more
surprising is the finding that the
mobility develops a minimum at $T/\Omega \lesssim 1$ and that the high-temperature limit is
recovered only after
 a maximum at $T/\Omega \sim \alpha$, see Fig.~\ref{fig:3}.
The $\mu M T$ value at the minimum is about $1/30$.

The mobility minimum for lattice polarons is a well-established phenomenon
(see, {\it e.g.},
Refs.~\cite{Holstein59,Friedman63,minimum,minimum1,minimum2}),
known as ``activated hopping" between
lattice sites. However, we are not aware of any work on the mobility minimum for
Fr\"{o}hlich polarons because the notion of
hopping is ill-defined in the continuum.
To understand the origin of the non-monotonic
behavior, we note that for $\alpha =6$ the effective mass renormalization
at low temperatures and small momenta is very strong, $M^{*}/M \approx 7.3$ \cite{us2},
while the physics at high temperatures is still captured by the Migdal theorem.
The mobility minimum then emerges from the competition between the decreasing number of
thermal excitations and the strong particle mass renormalization:
The former dominates at $T/\Omega\ll 1 $,  while the latter
is more important at $T/\Omega \gtrsim 1$.
\begin{figure}
\vspace{0.1cm}
\centering
\includegraphics[width=0.9\linewidth]{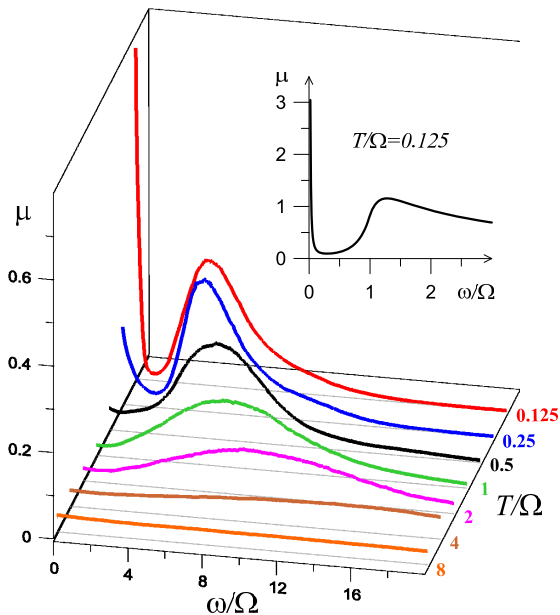}
\caption{\label{fig:4} (color online).
The real part of the
mobility for $\alpha = 6$
as a function of frequency
at different temperatures.
At low temperatures (red and blue curves) we observe a well-defined Drude peak.
In the temperature interval $ 0.5 \le T/\Omega \le 4$
the mobility
has a local
minimum at $\omega =0$ that cannot be explained within the kinetic-equation approach.
Only at $T/\Omega =8$ the situation changes back to having a (barely visible)
broad maximum at $\omega =0$. For comparison, the inset displays the real of part of the mobility
for a Holstein polaron at weak coupling \cite{Supplemental}.
}
\end{figure}

If
renormalization of the effective mass, which is temperature and energy dependent, does play
such an important role,
 then the dependence of $\mu$ on $\omega$
should be radically different from the prediction of the Drude model in Eq.~(\ref{Drude}).
To check this assertion we present the real part of $\mu(\omega,T)$ in Fig.~\ref{fig:4}.
At the two lowest temperatures ($T=0.125\, \Omega$ and $T=0.25\,\Omega$),
$\mu(\omega,T)$ exhibits a narrow Drude peak at low frequencies and a broad maximum at intermediate frequencies.
In the context of small lattice polarons,
the broad maximum in the mobility
is commonly attributed to
 ionization of bound states \cite{Emin}.
In a broader context, however, this maximum is referred to as ``Holstein band'' and arises because
emission of phonons is activated for $\omega$ above the characteristic phonon frequency \cite{Allen,Timusk}.
The Fr{\"o}hlich model is just one example; another one is the Holstein model the result for which is
shown in the  inset in Fig.~\ref{fig:4}.

For $T/\Omega \gtrsim 0.5$ the Drude peak in the mobility disappears. Moreover,
for $T/\Omega\gtrsim 2.0$
 the broad maximum becomes barely distinguishable.
Any attempt to extract the scattering time from the frequency dependence of the mobility
obviously fails in this regime. Our interpretation instead is to use the physically appealing
relation $\tau^{-1} \sim E \sim T$ in order to conclude that the effective mass defined through
$M^* \sim \tau/ \mu $ undergoes strong renormalization
as the temperature is lowered down to $T< \Omega$ (which is further confirmed by calculating
the effective mass at $T=0$).

Even if the mobility retains a Drude peak but the MIR criterion is violated, the scattering time can not be
extracted from the Drude formula.
This is supported by an example of a heavy particle with a large transport scattering
cross-section embedded into
 a Fermi sea,
{\it e.g.}, an electron bubble in ${}^3$He.
Its mobility
is described by Eq.~(\ref{Drude}), cast in the form $\mu (\omega) = 1/M(\Gamma -i \omega)$,
with a width $\Gamma \approx \text{const}$ down to temperatures exponentially lower than
$\Gamma$~\cite{ionsHe3}. This result is also well known in the context of Ohmic dissipative
models \cite{Schmid,Bulgadaev}. Interpreting $\Gamma$ as the scattering rate at temperatures
$T \ll \Gamma$ implies that the particle mass is not renormalized. However,
if the Planckian bound on the scattering rate
$1/\tau (T)$ is to be obeyed, one is forced to conclude that the effective mass has to diverge as
$M^{*} \propto M \Gamma /T$ in order to maintain a constant DC mobility $\mu = 1/M\Gamma \sim \tau(T) /M^{*}(T)$.
That the last interpretation is physically correct is revealed
by considering the superfluid state at $T \ll T_C \ll \Gamma$,
when the undamped particle motion is controlled by the strongly renormalized effective mass
$M^{*} \propto M \Gamma /T_C$ \cite{ionsHe3}.

In conclusion, we have studied the problem of electron mobility in ionic
semiconductors in the temperature regime where the
MIR
criterion for the
applicability of the kinetic-equation approach is violated. We found that the
mobility can be orders of magnitude smaller than the
MIR value of $1/MT$.
This result is consistent with recently observed
MIR violation
for non-degenerate charge carriers in doped SrTiO$_3$ \cite{Behnia}.
At strong coupling, the mobility has a minimum at a temperature comparable to
the optical mode frequency and increases with
a further increase of the
 temperature
 despite
 the fact
that the number of thermally
excited phonons
grows linearly with $T$. We
ascribe
this behavior
to the ``undressing" of polarons
at higher temperatures.
After going through a maximum at $T \gg \Omega$, the mobility follows the $\mu \sim T^{-1/2}$
law predicted by the kinetic equation.

{\it Acknowledgments.}
We acknowledge stimulating discussions with K. Behnia and T. Timusk.
This work was initiated and partially performed at Aspen Center for Physics supported by the
National Science Foundation grant PHY-1607611. We acknowledge support
by the Simons Collaboration on the Many Electron Problem (N.P.), the National Science Foundation
under grant DMR-1720465 (N.V.P.) and DMR-1720816 (A.K. and D.L.M.),
the ImPACT Program of the Council for Science, Technology and Innovation (Cabinet Office, Government of Japan)
and JST CREST Grant Number JPMJCR1874, Japan (A.S.M.),
and by FP7/ERC Consolidator grant QSIMCORR No. 771891, the Nanosystems Initiative Munich and the Munich Center for Quantum Science and Technology (L.P.).
Open Data for this project can be found at \url{https://gitlab.lrz.de/QSIMCORR/mobilityfroehlich}

\end{document}




\title{Supplemental material for ''Polaron mobility in the ``beyond quasiparticles" regime" }
\maketitle

\subsection{Mobility in the high-temperature limit $T\gg \Omega$}
In the  high-temperature limit one can ignore vertex corrections thanks to the Migdal theorem
and use the kinetic equation (KE) to find the mobility. The KE reads (as in the main text we set $e$, $\hbar$, and $k_B$ to unity)
\begin{eqnarray}
{\mathbf v}_{\mathbf k}\cdot {\bf E} f'_{0{\mathbf k}}&=&2\pi \int \frac{d^3q}{(2\pi)^3} |V(q)|^2\left\{\delta(\varepsilon_{{\mathbf k}+{\mathbf q}}-\varepsilon_{\mathbf k}+\Omega)\left[N_{\bf q}( f_{{\mathbf k}+{\mathbf q}}-f_{\mathbf k})-f_{\mathbf k}(1-f_{{\mathbf k}+{\mathbf q}})\right]\right. \nonumber   \\
&&\left.+\delta(\varepsilon_{{\mathbf k}+{\mathbf q}}-\varepsilon_{\mathbf k}-\Omega)\left[N_{\bf q}( f_{{\mathbf k}+{\mathbf q}}-f_{\mathbf k})+f_{{\mathbf k}+{\mathbf q}}(1-f_{{\mathbf k}})\right]\right\}\,,
\end{eqnarray}
where $V({\bf q})=i\sqrt{2^{3/2}\pi\alpha\Omega^{3/2}/M^{1/2}q^2}$ is the matrix element of Fr{\"o}hlich electron-phonon interaction, $f_{\mathbf k}$ is a non-equilibrium distribution function,
$f_{0{\mathbf k}}$ is the Fermi function, $N_{\bf q}$ is the Bose function,
$\varepsilon_{\mathbf k}=k^2/2M$, $\Omega$ is the optical phonon frequency, and  ${\mathbf v}_{\mathbf k} = {\mathbf k}/M$.
For $T\gg \Omega$, $N_{\bf q}\approx T/\Omega \gg 1$ and all terms in the collision integral which do not contain $N_{\bf q}$ can be dropped. Because scattering is (quasi)elastic, we can also drop $\Omega$ in the $\delta$-functions.
After this, the KE acquires the same form as for impurity scattering
\begin{equation}
\left({\mathbf v}_{\mathbf k}\cdot {\bf E}\right) f'_{0{\mathbf k}}=4\pi \frac{T}{\Omega} \int \frac{d^3q}{(2\pi)^3} |V({\bf q})|^2\delta(\varepsilon_{{\mathbf k}+{\mathbf q}}-\varepsilon_{\mathbf k})( f_{{\mathbf k}+{\mathbf q}}-f_{\mathbf k})\,,
\end{equation}
which is solved by the Ansatz:
\begin{equation}
f_{\mathbf k}=f_{0{\mathbf k}}-\tau\left({\mathbf v}_{\mathbf k}\cdot {\bf E}\right)f'_{0{\mathbf k}}\, ,
\end{equation}
where $\tau$ is a function of $|{\mathbf k}|$. Solving for $\tau$, we obtain
\begin{equation}
\tau^{-1}=4\pi \frac{T}{\Omega} \int \frac{d^3q}{(2\pi)^3} |V({\bf q})|^2\delta(\varepsilon_{{\mathbf k}+{\mathbf q}}-\varepsilon_{\mathbf k})(1-\cos \hat{{\mathbf k}{\mathbf q}} ) \,,
\end{equation}
were $\hat{{\mathbf k}{\mathbf q}}$ is the angle between ${\mathbf k}+{\bf q}$ and ${\mathbf k}$.
(Note that the Coulomb singularity in $|V({\bf q})|^2$ is cancelled by the transport factor,
$1-\cos \hat{{\mathbf k}{\mathbf q}} = q^2/2k^2$.)
Elementary integration yields
\begin{equation}
\tau=\frac{1}{2\alpha} \frac{v_{\mathbf k}}{u_0}\frac{1}{T} \, ,
\end{equation}
where $u_0=\sqrt{2\Omega/M}$. Assuming that ${\bf E}$ is oriented along the $z$-axis, we obtain the mobility as
\begin{equation}
\mu =
\frac{1}{n}
 \int \frac{d^3k}{(2\pi)^3} (-f'_{0{\mathbf k}})\tau v^2_z =\frac{ e^2}{6\alpha Tu_0 n} \int^\infty_0 d\varepsilon \nu(\varepsilon) (-f'_{0{\mathbf k}}) v_{\mathbf k}^3 \,,
\end{equation}
where $n$ is the number density and $\nu(\varepsilon)=M^{3/2}\sqrt{\varepsilon}/\sqrt{2}\pi^2$ the density of states.
In the Maxwell-Boltzmann regime, \bea f_{0{\mathbf k}}= 
(2\pi/MT)^{3/2}n \exp(-\varepsilon/T),\label{MB}\eea and
an elementary integration leads to the final result
\begin{equation}
\mu = \frac{4}{3\sqrt{\pi}\alpha} \frac{1}{M} \frac{1}{\sqrt{\Omega T}} \,.
\end{equation}
\subsection{Frequency-dependent mobility in the Holstein model}
In this section, we illustrate how the maximum in the frequency dependence of the mobility (see Fig.~4 of the main text) can be understood 
within the weak-coupling theory of electron-phonon interaction. Such a maximum can be anticipated from fairly general grounds. Indeed, the real part of the Drude-like mobility [Eq.~(1) of the main text] is given by
\bea
\mu'(\omega,T)=\frac{\tau(\omega,T)}{M\left[\omega^2\tau^2(\omega,T)+1\right]},
\eea
where $\tau$ depends both on $\omega$ and $T$.  After passing a Drude peak of width $\sim 1/\tau(T,0)$, the mobility enters the ``collisionless'' regime in which 
\bea
\mu'(\omega,T)\approx \frac{1}{M\omega^2\tau(\omega,T)}.
\eea
For any kind of electron-phonon interaction, $1/\tau(\omega,T)$ increases with $\omega$ faster then $\omega^2$, and thus $\mu'(\omega,T)$ increases with $\omega$. This increase is curbed off at $\omega$ larger than the characteristic phonon frequency, when $1/\tau(\omega,T)$ becomes independent of $\omega$ and $\mu'(\omega,T)$ drops off with $\omega$. This is the origin of the maximum in $\mu'(\omega,T)$.

We illustrate this behavior on the example of Holstein model, \cite{Holstein59} which describes a  short-range electron-phonon interaction with optical phonons. 
To lowest order in the electron-phonon coupling, the imaginary part of self-energy 
is given by 
\bea
\Sigma''(\e)=-g^2\int \frac{d^3p}{(2\pi)^3}\int \frac{d\Omega'}{2\pi} G''_0(\e+\Omega',p) D''(\Omega')\left[n_F(\e+\Omega')+N_0(\Omega')\right],
\eea
where $g$ is the electron-phonon coupling constant,
\bea
G_0(\e,p)=\frac{1}{\e-\frac{p^2}{2M}+\mu+i0^+}
\eea
is the retarded Green's function of free electron gas with chemical potential $\mu$,
 \bea
D(\Omega')=\frac{2\Omega}{\Omega'^2-\Omega^2+i\text{sgn}\Omega' 0^+}
\eea
is the Green's function of optical phonons with frequency $\Omega$, $n_F(\e)=\left(e^{\e/T}+1\right)^{-1}$ is the Fermi function with $\mu=0$, and $N_0(\Omega')$ is the Bose function.

For non-degenerate electrons, it is convenient to introduce a new energy variable $E\equiv \e+\mu$, in terms of which
\bea
\Sigma''(E)=-\Sigma_0\left\{ \sqrt{\frac{E+\Omega}{\Omega}}\theta(E+\Omega)\left[N_0(\Omega)+f_0(E+\Omega)\right]+\theta(E-\Omega)\sqrt{\frac{E+\Omega}{\Omega}}\left[N_0(\Omega)+1-f_0(E-\Omega)\right]\right\}.\label{SE}
\eea
Here, $\Sigma_0=\pi g^2\nu(\Omega)/2$, $f_0(E)$ is the Fermi function in the Maxwell-Boltzmann limit given by Eq.~(\ref{MB}),  
$\nu(E)$ is the electron density of states, and $\theta(x)$ is the Heaviside step-function.  For a short-range electron-phonon interaction, the vertex corrections to the current-current correlation function are absent,  and the Kubo formula for the real part of the mobility to leading order in $g$ is reduced to 
\bea
\mu'(\omega)=\frac{1}{6\pi^2 M^2\omega n}\int^\infty_{-\infty}dE \int_0^\infty dp p^4 G''(E+\omega,p)G''(E,p)\left[f_0(E)-f_0(E+\omega)\right],
\label{Kubo1}
\eea
where $n$ is the number density of electrons, $M$ is the electron mass, \bea
G(E,p)=\frac{1}{E-\frac{p^2}{2M}-i\Sigma''(E)},
\eea
and $\omega$ is chosen to be positive without loss of generality.
At weak coupling, $|\Sigma''(E)|\ll E$. The spectral functions in Eq.~(\ref{Kubo1}) are sharply peaked only for $E>0$, which effectively limits the range of integration over $E$ to interval $(0,\infty)$.
 Solving the integral over $p$ with condition $|\Sigma''(E)|\ll E$ taken into account and using  Eq.~(\ref{MB}) for $f_0(E)$, we obtain
 \bea
 \mu'(\omega)=-\frac{1}{M}\frac{2}{3\sqrt{\pi}\omega}\int^\infty_0 dE&& \frac{E^{3/2}+(E+\omega)^{3/2}}{T^{3/2}}e^{-E/T}\left(1-e^{-\omega/T}\right)\nn\\
 &&\times\frac{\Sigma''(E)+\Sigma''(E+\omega)}{\omega^2+\left[\Sigma''(E+\omega)+\Sigma''(E)\right]^2}.
 \eea
The remaining integral over $E$ is solved numerically. The resultant frequency dependence $\mu$ is shown in the inset of Fig.~4 of the main text for $T/\Omega=0.125$ and $\Sigma_0/\Omega=0.5$.

\begin{figure}[!]
\vspace*{0.0cm}
\includegraphics[scale=0.40,angle=0,width=0.88\columnwidth]{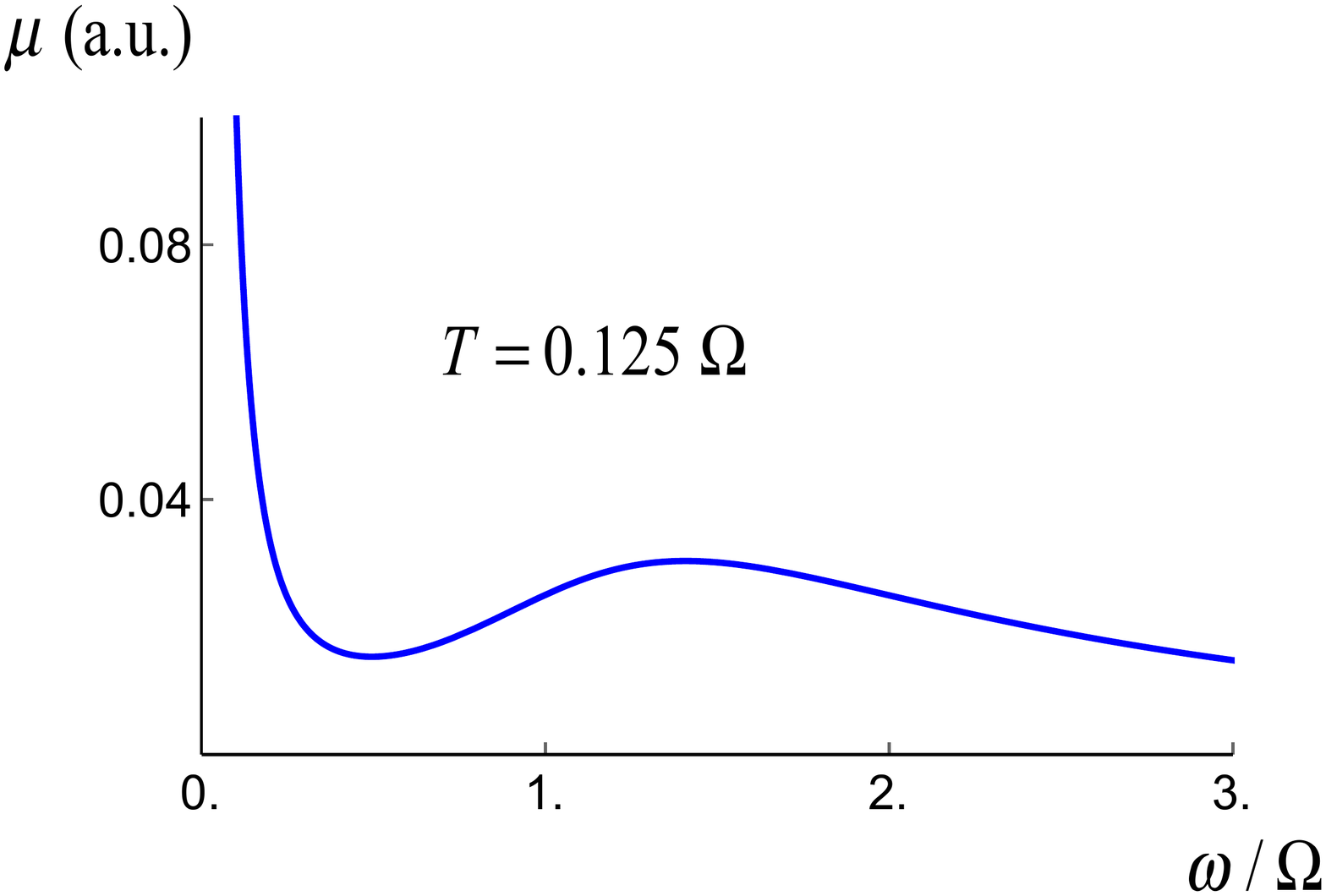}
\vspace*{0.0cm}
\caption{
\label{fig_deg} (color online) Frequency dependence of the real part of the mobility in the Holstein model for degenerate electron gas. The prefactor of the self-energy in Eq.~(\ref{SE1}) is chosen as $\tilde\Sigma_0=0.1\Omega$.}
\end{figure}

For non-degenerate electrons ($T\ll \mu\approx E_F$, where $E_F$ is the Fermi energy), one only has to replace $E$ by $E_F$ in the square-root factors  and by $\e$ in the Fermi functions in Eq.~(\ref{SE}). Assuming also the adiabatic limit ($\Omega\ll E_F$), we then obtain for the imaginary part of the self-energy
\bea
\tilde\Sigma''(\e)=-\tilde\Sigma_0\left\{\theta(\e+\Omega)\left[N_0(\Omega)+f_0(\e+\Omega)\right]+\theta(\e-\Omega)\left[N_0(\Omega)+1-n_F(\e-\Omega)\right]\right\},\label{SE1}
\eea
where $\tilde\Sigma_0=\pi g^2\nu(E_F)/2$. In the Kubo formula, one replaces the integration measure $d^3p/(2\pi)^3$ by $\nu(E_F) d\xi_p d\mathcal{O}/4\pi$, where $\mathcal{O}$ is the solid angle and $\xi_p=p^2/2m-E_F$, and takes the factor of $p/M\approx v_F\equiv\sqrt{2E_F/M}$ outside the integral. After integration over $\xi_p$, we then find
\bea
\mu(\omega)=-\frac{1}{M\omega}\int^\infty_{-\infty} d\e\left[n_F(\e)-n_F(\e+\omega)\right] \frac{\Sigma''(\e)+\Sigma''(\e+\omega)}{\omega^2+\left[\Sigma''(\e+\omega)+\Sigma''(\e)\right]^2}.
\eea
 The resultant $\mu(\omega)$ is shown in Fig.~\ref{fig_deg}
for $T/\Omega=0.125$ and $\tilde\Sigma_0/\Omega=0.1$.


\subsection{Analytic continuation}

To extract the mobility and its uncertainty from solutions of Eq.~(10) in the main text we employed the
stochastic optimization with consistent constraints (SOCC) method \cite{us2,us3}.
First, we find $N$ solutions,  $\left\{ \mu_i(\omega) \right\}$, $i=1, \ldots, N$,
that would reproduce Matsubara frequency data within their error bars (our accuracy was
better than five significant digits for the lowest frequencies).
The set of mobilities $\left\{ \mu_i \right\}$ 
was obtained from the limiting values of frequency-dependent solutions,
$\mu_i =  \mu_i(\omega \to 0)$, for which we considered several protocols:
(i) The value of $ \mu_i(\omega)$ at the lowest frequency, (ii) linear, and (iii) quadratic
fits+extrapolation of $\mu_i(\omega)$ data to $\omega = 0$ (from the range
of small frequencies where $\mu_i(\omega)$ is structureless).
However, differences in final results between these protocols were found to be much
smaller than the statistical error bars described below.
\begin{figure}[htbp]
\vspace*{0.0cm}
\includegraphics[scale=0.40,angle=0,width=0.88\columnwidth]{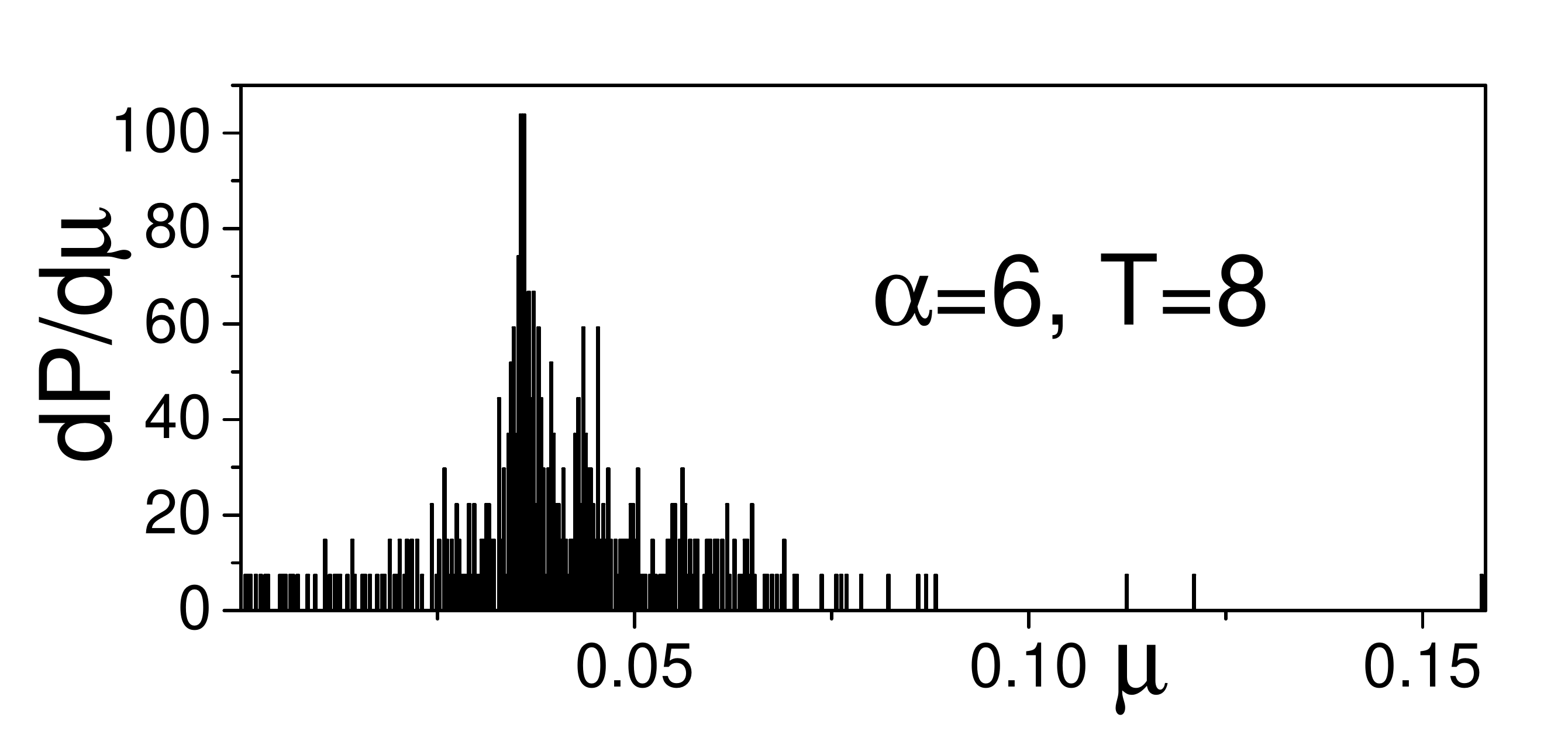}
\vspace*{0.0cm}
\caption{
\label{fig_muhist} Typical probability density distribution of $\mu_i$ values.
}
\end{figure}

Despite having highly accurate data in the Matsubara frequency domain the distribution
of possible mobility values turns out to be rather broad, see Fig.~\ref{fig_muhist}, which is a typical scenario for such problems.
Given a typical probability distribution as shown in  Fig.~\ref{fig_muhist} we implemented
three alternative protocols for presenting the final answer for $\mu$ and its error bars: \\
(i) An unrestricted average over all mobility values, 
$\mu = \bar{\mu} = (1/N) \sum_{i=1}^N \mu_i$; \\
(ii) An average over a restricted set of $N^*$ solutions with $\mu_i \ne 0$ 
(some solutions have zero spectral density at $\omega=0$ and they 
are accounted for by the SOOC method),
$\mu=\bar{\mu}^{\rm (r)} = (1/N^*) \sum_{i=1}^{N} \mu_i \Theta (\mu_i +0^+)$; \\
(iii) The most probable value; {\it i.e.}, the location of the maximum of the probability distribution
function shown in Fig.~\ref{fig_muhist}. We denote it as $\mu = \mu^{\rm (max)}$.

Since the distribution of mobilities is asymmetric (limited from below and unlimited from above) 
we estimate both the upper and lower error bars.
The upper(lower) error bars are then defined though asymmetric dispersions computed from statistics 
of $N_>$ ($N_<$) solutions with $\mu_i > \mu$ ($\mu_i < \mu$):
\begin{equation}
\sigma^{\rm (u)} = \sqrt{
\frac{1}{N_>} \sum_{i=1}^{N_>} (\mu_i-\mu)^2 \Theta(\mu_i-\mu )} \; , \;\;\;\;
\sigma^{\rm (d)} = \sqrt{
\frac{1}{N_<} \sum_{i=1}^{N_<} (\mu_i-\mu)^2 \Theta(\mu-\mu_i)} \; .
\label{dmu}
\end{equation}
We find that in all cases considered the error bars
$\sigma^{\rm (u)}$ and $\sigma^{\rm (d)}$ are much larger than differences 
between the alternative definitions of $\mu$. For example, for $\alpha=6$ and 
$T/\Omega=8$, see Fig.~\ref{fig_muhist}, the distribution is such that    
$\bar{\mu}=0.037$, $\bar{\mu}^{\rm (r)}=0.040$, $\mu^{\rm (max)}=0.036$, 
and $\sigma^{\rm (d)}=0.016$, $\sigma^{\rm (u)}=0.018$.
All data presented in the main text are based on averages over restricted sets of $N^*$ 
solutions; i.e. $\mu =\bar{\mu}^{\rm (r)}$ and   
\begin{equation}
\mu (\omega) = (1/N^*) \sum_{i=1}^{N} \mu_i(\omega) \Theta(\mu_i+0^+).
\label{reoc}
\end{equation}